%% file: main.tex
\documentclass[sigconf]{acmart}
\AtBeginDocument{%
  }

\copyrightyear{2025}
\acmYear{2025}
\setcopyright{rightsretained}
\acmConference[CHI EA '25]{Extended Abstracts of the CHI Conference on Human Factors in Computing Systems}{April 26-May 1, 2025}{Yokohama, Japan}
\acmBooktitle{Extended Abstracts of the CHI Conference on Human Factors in Computing Systems (CHI EA '25), April 26-May 1, 2025, Yokohama, Japan}\acmDOI{10.1145/3706599.3720219}
\acmISBN{979-8-4007-1395-8/2025/04}

\usepackage{subcaption}
\usepackage{multirow}

\newcommand{\vl}{\textsc{View Leader}}
\newcommand{\vf}{\textsc{View Follower}}

\newboolean{showrevisions}
\setboolean{showrevisions}{false}

\ifthenelse{\boolean{showrevisions}}{\definecolor{newcolor}{rgb}{0.15, 0.15, 1}
\definecolor{chcolor}{rgb}{0.15, 0.15, 1}}
{\definecolor{newcolor}{rgb}{0, 0, 0}
\definecolor{chcolor}{rgb}{0, 0, 0}}

\newcommand{\newtext}[1]{\textcolor{newcolor}{#1}}

\newcommand{\changetext}[1]{\textcolor{chcolor}{#1}}

\begin{document}

\title{Decoupled Hands: An Approach for Aligning Perspectives in Collaborative Mixed Reality}


\author{Matt Gottsacker}
\affiliation{%
  \institution{University of Central Florida}
  \city{Orlando}
  \state{Florida}
  \country{USA}}
\email{mattg@ucf.edu}

\author{Nels Numan}
\affiliation{%
  \institution{University College London}
  \city{London}
  \country{England}}
\email{nels.numan@ucl.ac.uk}

\author{Anthony Steed}
\affiliation{%
  \institution{University College London}
  \city{London}
  \country{England}}
\email{a.steed@ucl.ac.uk}

\author{Gerd Bruder}
\affiliation{%
  \institution{University of Central Florida}
  \city{Orlando}
  \state{Florida}
  \country{USA}}
\email{gerd.bruder@ucf.edu}

\author{Gregory F. Welch}
\affiliation{%
  \institution{University of Central Florida}
  \city{Orlando}
  \state{Florida}
  \country{USA}}
\email{welch@ucf.edu}

\author{Steven Feiner}
\affiliation{%
  \institution{Columbia University}
  \city{New York}
  \state{New York}
  \country{USA}}
\email{feiner@cs.columbia.edu}

\renewcommand{\shortauthors}{Gottsacker et al.}

\begin{abstract}
When collaborating relative to a shared 3D virtual object in mixed reality (MR), users may experience communication issues arising from differences in perspective.
These issues include occlusion (e.g., one user not being able to see what the other is referring to) and inefficient spatial references (e.g., ``to the left of this'' may be confusing when users are positioned opposite to each other).
This paper presents a novel technique for automatic perspective alignment in collaborative MR involving co-located interaction centered around a shared virtual object.
To align one user’s perspective on the object with a collaborator’s, a local copy of the object and any other virtual elements that reference it (e.g., the collaborator’s hands)
are dynamically transformed.
The technique does not require virtual travel and preserves face-to-face interaction.
We created a prototype application to demonstrate our technique and present an evaluation methodology for related MR collaboration and perspective alignment scenarios.
%
\end{abstract}



\newcommand{\subfigwidth}{0.25\linewidth}
\begin{teaserfigure}
    \centering
    \begin{subfigure}{\subfigwidth}
        \includegraphics[width=\linewidth]{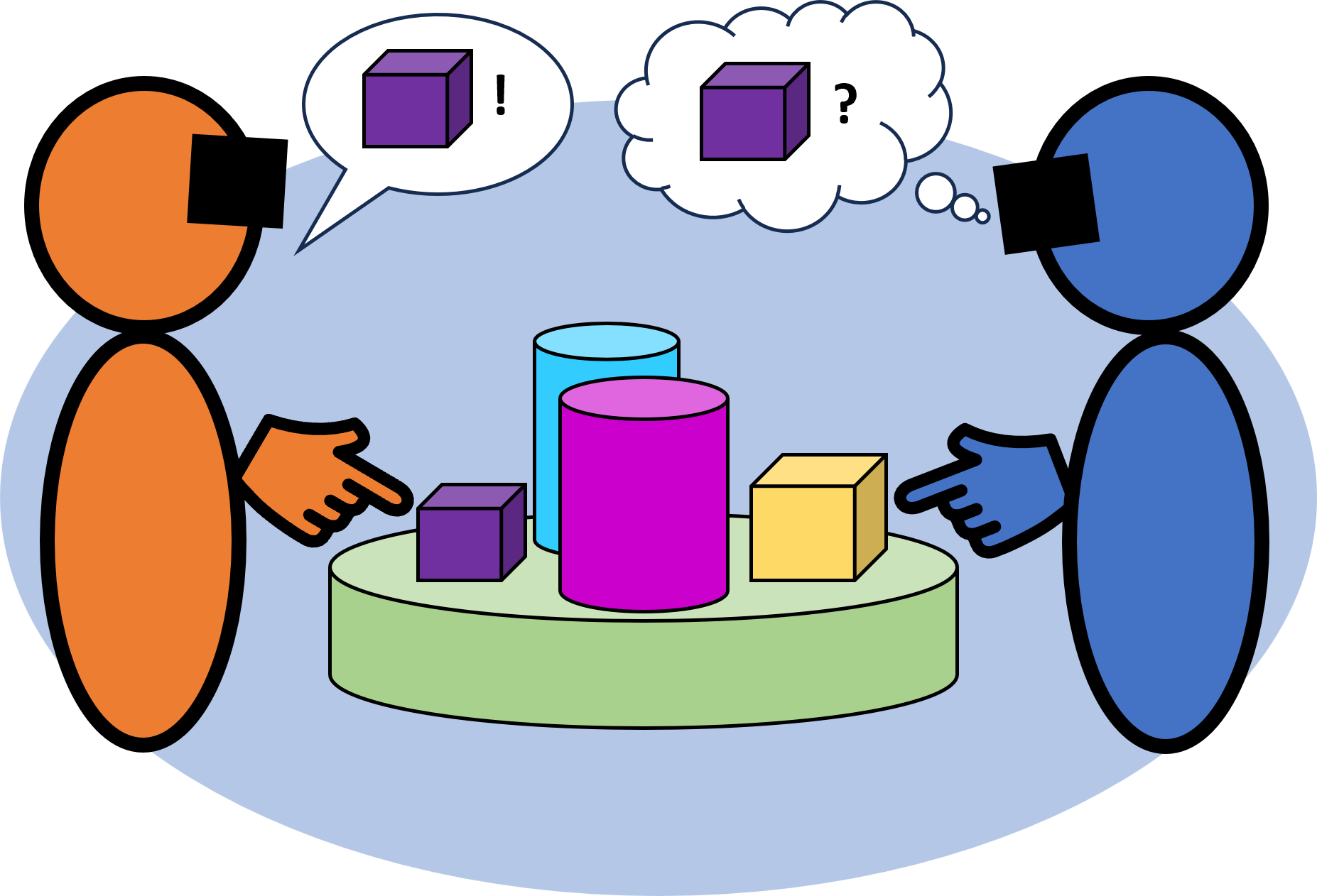}
        \caption{}
    \end{subfigure}
    \hspace{1em}
    \begin{subfigure}{\subfigwidth}
        \includegraphics[width=\linewidth]{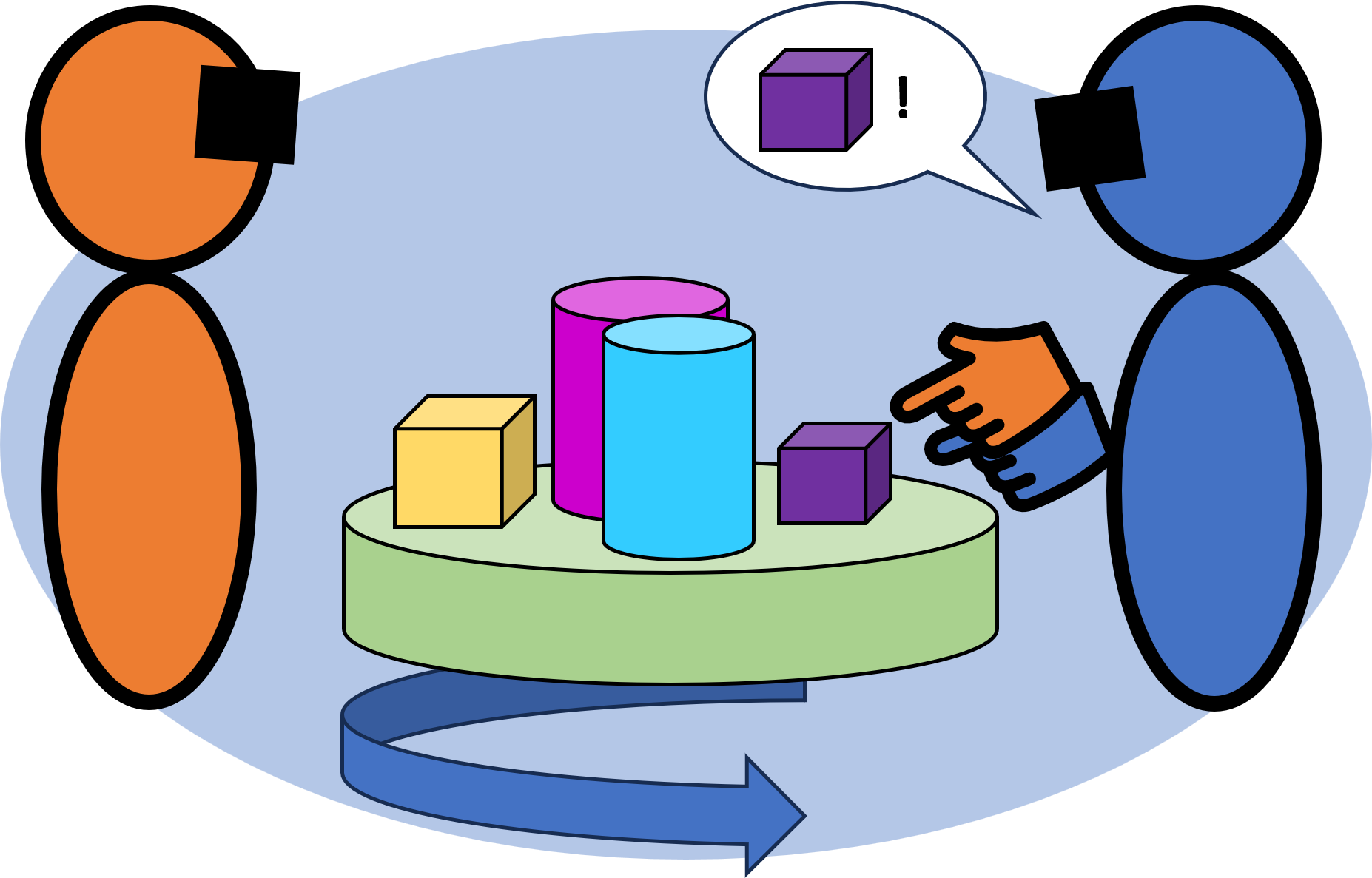}
        \caption{}
    \end{subfigure}
    \hspace{1em}
    \begin{subfigure}{\subfigwidth}
        \includegraphics[width=\linewidth]{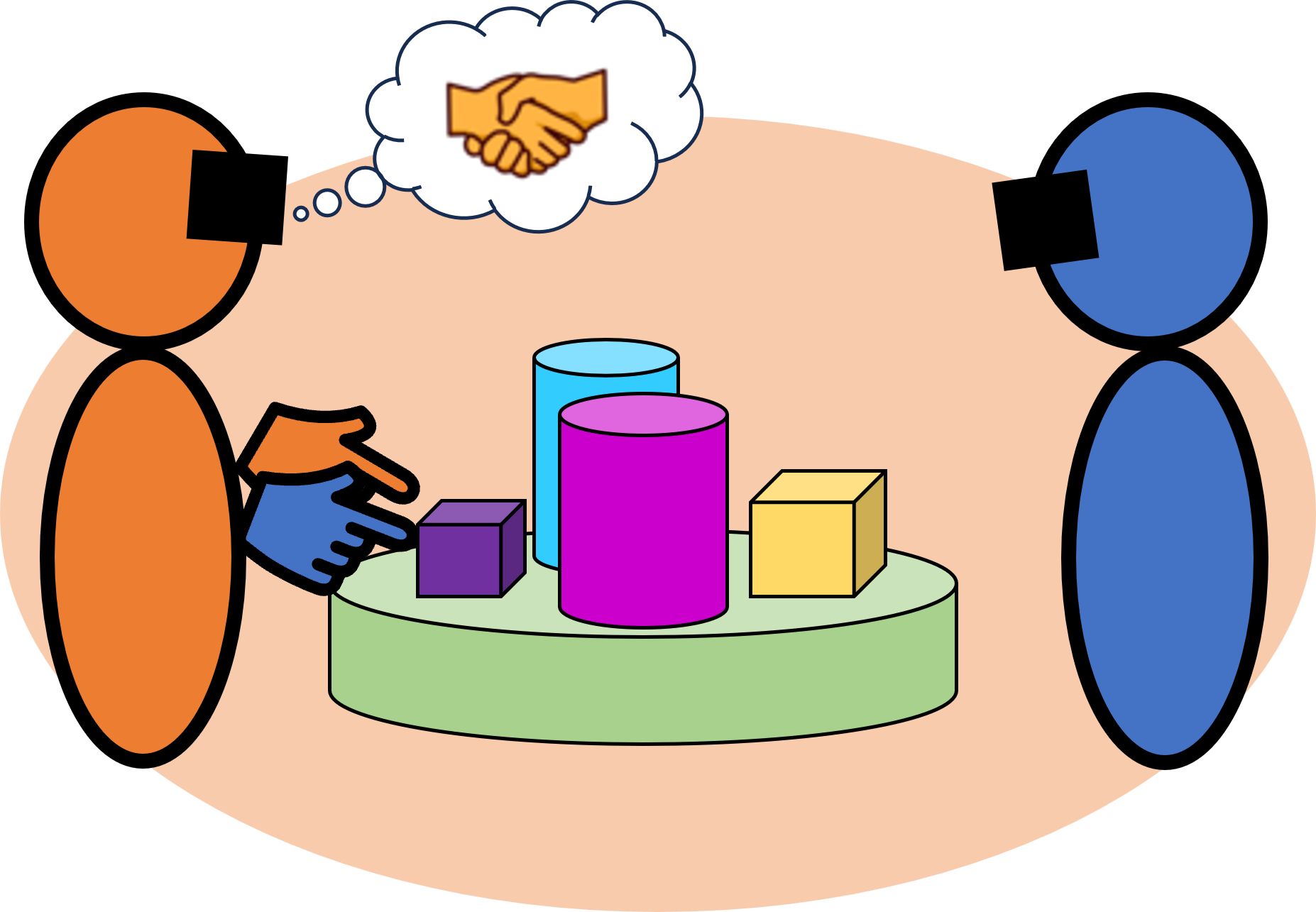}
        \caption{}
    \end{subfigure}
  \caption{
  (a) Mixed Reality users collaborating with shared virtual objects can encounter occlusion issues when objects block other objects from a collaborator's view.
  In our approach, each user has a local copy of the shared objects.
  To automatically align a user's perspective with their collaborator's, our technique (b) transforms the user's local objects so that the user has the same view as their collaborator.
  The virtual representations of the collaborator's hands transform with the objects.
  (c) The collaborator can see when the user is aligned with their perspective because the user's virtual hands inversely transform to reference the proper virtual position on their (un-rotated) local object.
  The proper positioning of the virtual hands for both users allows them to make intuitive references to the objects.
  }
  \Description{This figure illustrates a Mixed Reality collaboration scenario where two users, represented as orange and blue avatars, interact with shared virtual objects on a table. In panel (a), the users have different perspectives, leading to occlusion issues; the orange user sees the purple object clearly, while the blue user’s view is partially blocked. Panel (b) shows how the system aligns perspectives by transforming the blue user’s local copy of the objects to match the orange user’s perspective, with the blue user's virtual hands adjusting accordingly. Finally, in panel (c), the users share an aligned view where their virtual hands are correctly repositioned to reference the objects intuitively, reducing confusion and occlusion. The background colors in each panel indicate different perspective states, transitioning from a misaligned view to an adjustment phase and then to a fully aligned collaborative view.}
  \label{fig:teaser}
\end{teaserfigure}


\maketitle

\input{sections/01_intro}
\input{sections/02_background}

\input{sections/03_system}

\input{sections/04_evaluation}
\input{sections/05_conclusion}


\begin{acks}
This material includes work supported in part by the Office of Naval Research under Award Numbers N00014-21-1-2578 and N00014-21-1-2882 (Dr. Peter Squire, Code 34);
the AdventHealth Endowed Chair in Healthcare Simulation (Prof. Welch); DZYNE subaward DRP009-S-001 from DARPA;
National Science Foundation Award Number 2037101;
and the European Union’s Horizon 2020 Research and Innovation Programme under Grant Agreement No 739578.
\end{acks}

\bibliographystyle{ACM-Reference-Format}
\bibliography{sample-base}










\end{document}

%% file: sections/01_intro.tex
\section{Introduction}

In computer-supported cooperative work (CSCW), understanding collaborator attention is a fundamental aspect of workspace awareness and collaboration in digital systems~\cite{gutwin2002descriptive}.
In support of this, \emph{view sharing} is the process by which collaborators are able to see the same digital content simultaneously, and transfer control among each other.
Engelbart et al.~\cite{engelbart1968research} introduced perhaps the first example of view sharing in collaborative document editing applications in 1968, and this is now an integral element of a variety of digital collaboration applications.
For example, it is common for a user in a video conferencing meeting to share their screen when presenting a slideshow 
and use their mouse to direct others' attention to specific objects.
While view sharing in this way, the webcam footage of the users is also displayed, which preserves some social cues present in face-to-face interaction.

With co-located mixed reality (MR) setups in which multiple users collaborate using shared virtual objects (such as in table-based collaboration for construction planning~\cite{agrawala1997two, spindler2012table}, scientific visualization~\cite{agrawala1997two, obeysekare1996workbench, spindler2012table, yu2022duplicated}, and geospatial visualization and planning~\cite{agrawala1997two}),
maintaining strong workspace awareness is important.
In these scenarios, collaborators can experience mismatched viewpoints, where occupying different positions in the physical and/or virtual worlds results in different perspectives of the same virtual object. 
Researchers have shown that such asymmetry in perspectives and interaction capabilities can make it difficult to understand each other's spatial references~\cite{chastine2007referencing} and lead to communication issues~\cite{oda20123d}.
For example, perspective differences can make it difficult to refer to specific features or regions of a virtual object, as a gesture or verbal reference may not be well-understood across viewpoints. 
In addition, certain parts of the virtual content or a collaborator’s actions (e.g., pointing or gesturing) can be occluded depending on one’s position. 

To explore view-sharing interfaces in co-located MR scenarios, we consider the 
simple
scenario of a single \vl{} sharing their view with a single \vf{}.
Note that ``leader'' and ``follower'' may have nothing to do with the actual roles of the users collaborating (although it might; co-located immersive presentation~\cite{gottsacker2023presentation, gottsacker2025pres} and education~\cite{drey2022collaborative} scenarios commonly involve one user primarily directing the experience). 
Here, we use these terms merely to help differentiate between the users.
The most straightforward approach to accomplish this view sharing is to reproduce the \vl{}'s spatial viewpoint for the \vf{} in such a way that both collaborators see identical virtual stereo imagery.
However, this approach can present multiple challenges.
In this scenario, the \vl{} controls the \vf{}'s virtual position and rotation, leading the \vf{} to experience uncontrolled optic flow translations and rotations, which can cause cybersickness~\cite{chang2020virtual}. 
Furthermore, if the \vf{} were to occupy the same virtual position as the \vl{}, they would be unable to see the \vl{}'s face, hindering understanding of social cues.

In this work, we present an approach for MR view sharing designed to account for all of these challenges and provide intuitive communication and group awareness while preserving face-to-face interaction cues.
We propose a perspective alignment technique that provides each collaborator with their own copy of the virtual objects on which they are collaborating.
Rather than automatically translating and rotating the \vf{} through the \emph{environment}, which causes the aforementioned issues, 
the \vf{}'s 
copies of the \emph{objects} are translated and rotated to achieve alignment.
The same transformation is applied to virtual representations of the \vl{}'s hands so that when the \vl{} points to a virtual object, their virtual hand will be displayed to the \vf{} in the proper virtual location.
This idea can be applied to aligning perspectives with differences in scales, such as when the \vl{} increases the size of the object or decreases the size of their view frustum to achieve a more immersive view,
but that kind of transitional interface is beyond the scope of the work presented in this paper.

%% file: sections/02_background.tex
\section{Background}
\label{sec:background}
Our work is inspired by and grounded in research on group awareness and novel techniques for view sharing in collaborative MR environments.

\subsection{Group Awareness in Collaborative MR}

In 2002, Gutwin \& Greenberg~\cite{gutwin2002descriptive} developed a framework identifying the primary elements of workspace awareness that a multi-user system should support in order to provide a good collaborative experience.
At any given moment, collaborators should be aware of \textit{who} is involved in the workspace, \textit{what} they are working on, and \textit{where} they are working or attending to~\cite{gutwin2002descriptive}.
\newtext{
While this framework was initially focused on remote collaboration scenarios,
more recently 
it has been re-contextualized and examined in research on co-located MR collaboration, which has shown that there are substantial challenges to providing high workspace awareness even when collaborators share the same physical space~\cite{radu2021survey}.}
\changetext{
For example,
researchers have pointed out that interfaces should be designed such that collaborators can understand and direct each other's attention spatially (in 3D) and efficiently~\cite{chastine2007referencing, radu2021survey, oda20123d}.}
Challenges to maintaining this shared workspace awareness in MR arise due to 
occlusion (i.e., when one user is viewing things that are blocked from another user's perspective).
Additionally, seeing the workspace from different angles can result in less efficient communication if users do not easily understand references and need to verbally clarify a collaborator's references~\cite{kiyokawa2002communication}.
\newtext{
Differing perspectives can also lead to the users spending additional time manipulating the workspace or physically moving through the 
environment, to better understand each other~\cite{chastine2007referencing}.}
To overcome these issues, perspective alignment interfaces should be designed so that users can easily and quickly achieve common ground for spatial references.




\changetext{
Researchers have emphasized the importance of collaborators' abilities to observe each other's facial expressions and body language, which are important for making collaborators aware of each other's emotional states~\cite{radu2021survey} and building trust~\cite{ishii1994iterative}.
For this reason, we chose to include an MR-based face-to-face capability.
}
It should be noted, however, that there are limitations imposed by current consumer MR head-worn displays (HWDs).
For one, an HWD covers the upper half of a user's face, which reduces the observability of their facial expressions and can hamper social interaction~\cite{mcatamney2006examination, gottsacker2025pres}.
Researchers have proposed various methods to restore these cues (e.g., by displaying the users' eyes on the outside of the HWD~\cite{bozgeyikli2024googly, chan2017frontface, mai2017transparent, matsuda2021reverse} or sharing other invisible cues about the user's physiological or mental states to enhance communication~\cite{dey2017physiological, gottsacker2022cues}).

Our work seeks to support efficient spatial referencing through aligned perspectives while preserving the social interaction cues of face-to-face interaction in co-located collaborative MR.


\subsection{Mixed Reality View Sharing Techniques}

\changetext{
Researchers have explored a variety of techniques for aligning perspectives and supporting workspace awareness in collaborative MR scenarios where users have different viewpoints of shared virtual objects.}
Tserenchimed et al.~\cite{tserenchimed2024sharing} presented a technique to exactly align a \vf{}'s view with a \vl{} in a remote collaboration scenario involving a mobile AR user and a VR user.
Once the collaborators' perspectives are aligned, each user cannot see the other user's avatar, but they can see 3D rays cast from each other's controllers.
They found that their technique led to lower mental demand and faster task completion on a collaborative engine-fixing task.
Similarly, Le Ch{\'e}n{\'e}chal et al. developed the Vishnu system~\cite{le2016vishnu}, which positioned a remote expert using VR in the same virtual position as an AR user.
The system showed the expert's virtual arms emanating from the same place as the AR user's own arms in their view to point to things in the workspace.

Researchers have also explored re-mapping users' bodies and/or environments to provide aligned perspectives in collaboration.
For example, Congdon et al.~\cite{congdon2018merging} presented a method for distorting users' asymmetric virtual environments to provide a shared interaction space.
Additionally, in face-to-face scenarios, Sousa et al.~\cite{sousa2019negative} and Fidalgo et al.~\cite{fidalgo2023magic} aligned collaborator perspectives exactly, and then warped each collaborator's original hands and arms around the virtual object to point in the approximately correct virtual location for the other user.
However, these techniques were designed for and evaluated in environments in which the collaborators were interacting around a small tabletop with a virtual object placed on top of it.
Extending these methods to larger virtual objects would involve significantly warping the appearance of collaborators' original arms or hands.
Sim{\~o}es et al.~\cite{simoes2024sparc} took this approach in their SPARC system, which also re-targeted collaborator avatars. 
In this work, \vf{} avatars were re-targeted from the \vl{}'s perspective to prioritize face-to-face interaction (e.g., by positioning collaborators avatars across from the \vl{} rather than next to them).
When a \vf{} pointed to something, their arms and hands were then stretched and manipulated to point in the proper virtual location.
Hoppe et al. developed the ShiSha system for collaborative virtual reality~\cite{hoppe2021shisha}, which positioned \vf{}s in the same virtual location around a \vl{} to provide all users with aligned perspectives.
In the \vl{}'s view, however, the \vf{}s were positioned slightly off to the side so that they could still engage in face-to-face interaction.
The \vf{}s' avatars were then re-targeted to point to the approximately correct virtual location.

Our work presents an approach for aligning perspectives exactly (similar to other works such as~\cite{hoppe2021shisha, simoes2024sparc, le2016vishnu, sousa2019negative, fidalgo2023magic, tserenchimed2024sharing}) and that also preserves face-to-face interaction (similar to~\cite{simoes2024sparc, fidalgo2023magic, sousa2019negative}) but without warping users' avatars, \newtext{which potentially provides advantages for the scalability of the technique.}

%% file: sections/03_system.tex
\begin{figure*}
    \centering
    \includegraphics[width=1\linewidth]{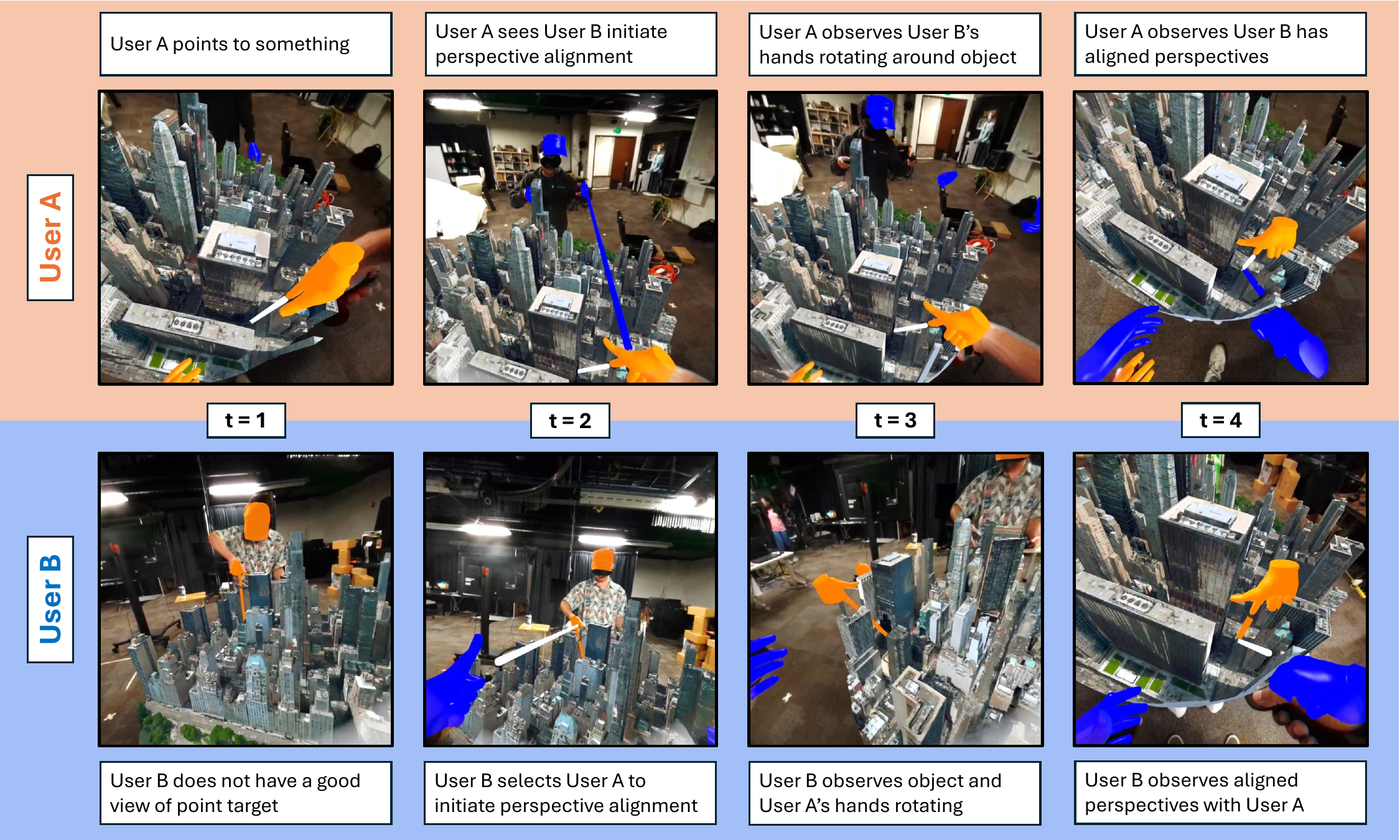}
    \caption{Screenshots from our collaborative AR system in use showing different stages of the perspective alignment process.
    At time \textbf{t = 1}, User A (top panel, orange) points at something that is out of view for User B (bottom panel, blue).
    At \textbf{t = 2}, User B points at User A and presses the trigger to activate a perspective alignment.
    At \textbf{t = 3}, the perspective alignment is in progress: User A can see User B's virtual hands rotating around the virtual object toward them; User B can see the virtual object and User A's virtual hands rotating toward them.
    At \textbf{t =4}, User A has become the \vl{} and User B has become the \vf{}. Specifically, User B's 3D map has rotated such that he can see the map from the same angle as User A. 
    Each user can see virtual representations of the other user's hands floating over the map in the correct virtual positions relative to the other user's view.}
    \Description{This figure consists of two rows of screenshots showing different stages of the perspective alignment process in a collaborative AR system. The top row represents the view from User A (orange avatar), while the bottom row shows the corresponding view from User B (blue avatar). Each column corresponds to a different time step (t = 1 to t = 4).
    At t = 1, User A points to an object of interest, which is out of view for User B. In the bottom panel, User B's perspective does not provide a clear view of the target object. At t = 2, User B selects User A to initiate a perspective alignment process, triggering a transformation of their virtual viewpoint. At t = 3, the alignment is in progress, and User A can see User B’s virtual hands rotating around the object. Meanwhile, User B observes the virtual object shifting and sees User A’s hands rotating in their view. By t = 4, the alignment is complete—User A has taken the role of the View Leader, while User B has become the View Follower. User B’s 3D map has rotated to match User A’s perspective, allowing them to see the scene from the same angle. Both users now see virtual representations of each other's hands floating over the map in the correct virtual positions relative to the other’s view.}
    \label{fig:screenshots}
\end{figure*}

\section{Perspective Alignment Technique Design \& Implementation}

This section describes the design principles that guided the creation of our Decoupled Hands perspective alignment technique, technical details of our approach, and the functional prototype we created to demonstrate it.

\subsection{Perpective Alignment Design Principles}

Similar to work on co-located MR presentations~\cite{gottsacker2025pres}, we based the design of our MR perspective alignment technique on 
high-level design goals that Kumaravel et al.~\cite{kumaravel2020trasceivr} developed from formative interviews about asymmetric interactions between users of VR and of non-immersive displays.
While Kumaravel et al.'s VR--desktop collaboration scenario was different from our MR scenario, their study of how disparate perspectives affect collaboration aligns with our research goals.

First, we aimed to support \textbf{independent exploration} for both users~\cite{kumaravel2020trasceivr} by enabling both users to explore and interact with the shared virtual content.
We also aimed to support \textbf{efficient and direct spatial references} to the virtual content~\cite{kumaravel2020trasceivr, chastine2007referencing, radu2021survey}, which is required to enable the collaborators to discuss specific aspects of the objects.
We provided virtual hand representations for each user and raycasts so they could clearly indicate specific points on an object to their partner.
Understanding these references is not an issue when both users have a good view of an object of interest.
However, when shared objects are large or complex, and it becomes difficult or time-consuming for users to obtain similar views, it is necessary to provide additional features to enable shared perspectives and support attention guidance tools for both users.
We also aimed to provide \textbf{stable virtual content} for both users~\cite{kumaravel2020trasceivr} to support their making precise references to content through pointing.
To achieve this principle, the perspective alignment on the virtual content should be triggered by the \vf{} and their view should not update continuously.

Last, we aimed to support \textbf{social interaction and co-presence} between users through social cues such as body language~\cite{kumaravel2020trasceivr, radu2021survey}.
Expressing and observing social cues is an important aspect of understanding others' emotional states~\cite{radu2021survey}, building trust~\cite{ishii1994iterative}, and establishing and gauging interest during computer-mediated interactions~\cite{murali2021affective}.
For this reason, we support face-to-face interaction through video see-through MR without modifying users' appearances.

\subsection{Perspective Alignment Technique}

Our approach provides each user with a local copy of the shared virtual objects around which the users are collaborating.
An axis of rotation is defined for the set of collaborative objects as the centroid of all objects in 3D space.
For table-based collaboration, this axis is set at the center of the table.
A \vf{} can align their perspective with a \vl{} by pointing their controller at them and pressing the trigger button.
The \vf{}'s copy of the virtual objects then rotates about their axis so that the \vf{} sees the objects from the same angle as the \vl{}.
The virtual objects always rotate around the shared objects' axis by the shortest angular distance between the two perspectives.

After it is computed once (i.e., when a \vf{} selects a \vl{}), this rotation value is used to transform all virtual copies of the \vl{}'s objects for the \vf{}.
The positions of the virtual copies are transformed around the work table by the same angular difference.
This method positions the \vl{}'s virtual hands in the exactly correct virtual position for the \vf{}.
In other words, if the \vf{}'s virtual object is rotated 100\textdegree\ with respect to the \vl{}'s object, the \vf{}'s virtual copy of the \vl{}'s hands will be offset from the \vl{} by the same 100\textdegree\ around the center of the virtual object.
The result is that when the \vf{} transforms their copy of the shared virtual object to match the \vl{}'s perspective, the \vf{} sees a virtual representation of the \vl{}'s hands floating in front of the \vf{}.
This decoupling of the \vl{}'s virtual hands from their original position allows the object-relative perspective alignment paradigm to work in a straightforward way (i.e., without any warping or re-targeting) with large virtual objects, large numbers of collaborators, and dynamic movements.

\begin{table*}[]
\begin{tabular}{ | p{12em} | p{10em} | p{25em} | }
\hline
\multicolumn{1}{|c|}{\textbf{Category}}                    & \multicolumn{1}{c|}{\textbf{Measure}} & \multicolumn{1}{c|}{\textbf{Description}}                 \\ \hline
\multirow{2}{12em}{\textbf{Task Performance}}                 & Speed                                 & How quickly users complete the spatial construction task. \\ \cline{2-3} 
                                          & Accuracy                                      & How accurately users fulfill the constraints of the sub-tasks.         \\ \hline
\multirow{4}{12em}{\textbf{User Experience}} & Cognitive load                            & NASA Raw Task Load Index~\cite{hart2006tlx}                                                   \\ \cline{2-3} 
                                          & Usability          & User Experience Questionnaire~\cite{laugwitz2008ueq}                                          \\ \cline{2-3} 
                                          & Social presence                     & Networked Minds Social Presence Inventory~\cite{harms2004internal}                              \\ \cline{2-3} 
                                          & Group awareness                     & Questions based on Gutwin \& Greenberg's workspace awareness framework~\cite{gutwin2002descriptive} \\ \hline
\multirow{3}{12em}{\textbf{Observed Collaboration Behaviors}} & Communication                         & Total word count, number of deictic phrases used          \\ \cline{2-3} 
                                          & Social interaction                            & How often participants look at each other                              \\ \cline{2-3} 
                                          & Perspective alignments & How often participants align perspectives                      \\ \hline
\end{tabular}
\caption{Planned evaluation metrics}
\label{tab:metrics}
\end{table*}

\subsection{Functional Prototype Application}

To demonstrate our approach, we developed an MR application that allows multiple users to interact with a 3D map and switch to each other's perspective.
This prototype was the result of a collaboration between two geographically separated research labs.
Screenshots of the system in action are shown in~\autoref{fig:screenshots}, and a video of the application in action is included in the supplementary materials.
In our demo scenario, one user can assume the role of \vl{} and teach the \vf{}s about features on the map (e.g., landmarks along a route).
The \vl{} can place virtual pins on the map and gesture to particular points of interest to communicate the map-based information.
The \vl{} has the best view of the map and its features, so the \vf{}s will need to align their perspectives with the \vl{}.

\subsubsection{Networking Implementation}

When users first connect to the application, they perform a calibration process to ensure the locally tracked positions of their devices are aligned in the same coordinate space on the networking server.


A virtual cap is placed on each user's head, and virtual hands mirror their physical hand movement.
A user can observe how closely these virtual objects match up with the physical head and hands of a collaborator to get a sense for how well their coordinate spaces are aligned.
They can recalibrate their coordinate spaces as needed.
Each user's virtual cap and hands are assigned a color to help users differentiate themselves from each other when aligning their perspectives.

\subsubsection{Prototype Hardware and Software}
We developed the application using Unity $2021.3.13$ and deployed the application to the Meta Quest Pro head-worn MR display.
The Quest Pro has a colorized video see-through mode that allows users to view their physical surroundings through the headset cameras, a key feature for supporting face-to-face collaborative MR.
The map data is streamed dynamically using the Microsoft Maps SDK for Unity\footnote{\url{https://github.com/microsoft/MapsSDK-Unity}}, which provides API access to detailed 3D terrain and building data for many cities around the world through Bing Maps.
The collaborators' head and hands transform data is networked using Photon Unity Networking 2\footnote{\url{https://www.photonengine.com/pun}}.

%% file: sections/04_evaluation.tex
\section{MR Object-Based Collaboration Evaluation Methodology}

While pilot tests using our demo application in our lab have shown promise, it is necessary to formally evaluate our technique.
In this way, this paper serves as a ``prequel'' to future research on the evaluation of this technique in collaborative MR scenarios.
Here, we introduce an experimental methodology to study the quality of collaboration around shared objects using different perspective alignment techniques in MR.
In a future user study, participant dyads will collaborate on a spatial task situated on a virtual table such as 
populating a 3D landscape with different buildings, objects, and features.
This 3D environment will be designed to include terrain and objects that can cause occlusion issues for users with different viewpoints.
Additionally, participants will be assigned sub-tasks that depend on the work of their collaborator and will be designed to encourage perspective sharing.
For example, one user will need to position a specific virtual object in the line-of-sight of an object that their collaborator is responsible for placing.
This task is inspired by earlier experiments on collaborative MR by Billinghurst et al.~\cite{billinghurst2002experiments}.

Different perspective alignment techniques (such as those described in ~\autoref{sec:background}, e.g., ~\cite{simoes2024sparc, fidalgo2023magic, hoppe2021shisha}) can be compared with our Decoupled Hands approach in trials that use different but similarly structured base environments and sub-tasks.
We will collect the measures listed in~\autoref{tab:metrics}, along with qualitative analysis of participant interviews, to gain insights into the impacts of the different perspective alignment techniques on the quality of participants' collaboration experiences.
We will measure task performance to assess the efficiency and effectiveness of users' collaboration (i.e., how well they coordinate their actions and accomplish shared goals).
On the user experience side, measuring users' perceptions of cognitive load and usability will provide insights into how intuitive and supportive users find a certain technique.
Additionally, users' perceptions of group awareness will indicate the degree to which a given technique allows users to make and understand each other's references.
Measuring social presence will assess how aware users feel of each other’s presence, which is crucial for establishing trust and engagement in MR collaboration.

Participants' behaviors are a useful data source as well.
For instance, tracking total word count and deictic phrase usage provides insight into the clarity and efficiency of verbal exchanges.
Measuring how often participants look at each other and make eye contact indicates the level of interpersonal engagement participants experience.
Last, assessing how often participants align their perspectives reveals how much participants relied on the technique.
Altogether, these measures provide insight into how well a given technique supports users' collaboration.

\section{Future Work}
While a study with just two users will reveal useful insights about the trade-offs between different kinds of perspective sharing techniques, we believe it is important to scale and test this technique in collaborations involving more than two users.
It will likely become more challenging for users to understand ``whose hands are whose'' when multiple users have triggered perspective alignments.
Tang et al.~\cite{tang2007videoarms} presented one approach for disambiguating collaborators' arms and hands in a surface-based mixed-presence system in which local users' hands were rendered semi-transparent and remote users' hands were opaque.
The color-matched virtual hands and caps in our application may work for this purpose up to a point, but larger numbers of collaborators may require additional visualization techniques such as labels or leader lines (even just temporarily) to disambiguate hand ownership.
Additionally, it may be necessary to incorporate view management techniques~\cite{bell2001view} such as automatically adjusting the position or transparency of visualizations and labels
to avoid users becoming overwhelmed or confused with the placement of collaborators' representations.

Another avenue for future work is to explore this technique for other physical objects in the collaborative environment.
By combining machine learning techniques for spatial understanding as well as object segmentation and classification (e.g., Augmented Object Intelligence~\cite{Dogan_2024_XRObjects}), our approach for transforming virtual objects relative to their 3D centroid can be extended to physical objects as well.
Such a system could create a virtual replica of relevant physical objects~\cite{oda2015replicas} and apply diminished reality techniques~\cite{cheng2022dr} to remove the physical objects from view.
This would allow the virtual replica objects to be transformed for each user in the same way as in the Decoupled Hands approach.

%% file: sections/05_conclusion.tex
\section{Conclusion}

This paper presents Decoupled Hands, a novel technique for achieving perspective alignment in co-located collaborative mixed reality environments involving a shared virtual object.
The technique efficiently aligns collaborators' perspectives on the object without requiring virtual travel and while supporting face-to-face interaction.
As it does not require warping or re-targeting the environment or collaborators inside it, our method has the potential to generalize to large virtual objects and large numbers of collaborators without sacrificing exact perspective alignment or unmediated face-to-face interaction.
We created a prototype system to demonstrate this technique involving a 3D map with real map data streamed from the Internet.